\documentclass[runningheads]{llncs}
\usepackage{graphicx}
\usepackage[pagebackref=true]{hyperref}
\begin{document}
\title{None Shall Pass: A blockchain-based federated identity management system}
\titlerunning{None Shall Pass}
%
\author{Shlok Gilda\inst{1}\orcidID{0000-0002-9355-4381} \and
Tanvi Jain\inst{1} \and
Aashish Dhalla\inst{1}}
\authorrunning{S. Gilda et al.}
%

\institute{Department of Computer and Information Science and Engineering, \\University of Florida, Gainesville FL 32608, USA\\
\email{\{shlokgilda, tjain, aashishdhalla\}@ufl.edu}}

\maketitle              
\begin{abstract}
Authentication and authorization of a user's identity are generally done by the service providers or identity providers. However, these centralized systems limit the user's control of their own identity and are prone to massive data leaks due to their centralized nature. We propose a blockchain-based identity management system to authenticate and authorize users using attribute-based access control policies and privacy-preserving algorithms and finally returning the control of a user's identity to the user. 

Our proposed system would use a private blockchain, which would store the re-certification events and data access and authorization requests for users' identities in a secure, verifiable manner, thus ensuring the integrity of the data. This paper suggests a mechanism to digitize documents such as passports, driving licenses, electricity bills, etc., issued by any government authority or other authority in an immutable and secure manner. The data owners are responsible for authenticating and propagating the users' identities as and when needed using the OpenID Connect protocol to enable single sign-on. We use advanced cryptographic algorithms to provide pseudonyms to the users, thus ensuring their privacy. These algorithms also ensure the auditability of transactions as and when required. Our proposed system helps in mitigating some of the issues in the recent privacy debates. The project finds its applications in citizen transfers, inter-country service providence, banks, ownership transfer, etc. The generic framework can also be extended to a consortium of banks, hospitals, etc.

\keywords{Blockchain \and Identity Management \and Authorization \and Authentication}
\end{abstract}

\section{Introduction}
\label{sec:introduction}
Identity providers or service providers act as third-party operators that authenticate and authorize the user's identity in an identity management system, and traditional systems are neither secure nor reliable. Password manager and single-sign-on provider, OneLogin was hacked, and all customers served by its US data center were affected~\footnote{https://www.zdnet.com/article/onelogin-hit-by-data-breached-exposing-sensitive-customer-data/}. Equifax, one of the most significant credit bureaus in the USA, was the victim of a data breach. Private records of approximately 148 million Americans were compromised in the breach, making it one of the largest cybercrimes related to identity theft~\cite{berghel2020equifax}. Even though these digital identities are portable, they are not secure. Several industries suffer from the problems of current identity management systems. The lack of interoperability between government departments and government levels takes a toll in the form of excessive bureaucracy. This, in turn, increases processes times and costs. In the field of education, there is a massive problem of fake academic certificates~\footnote{https://www.fox46.com/news/easy-to-get-fake-degrees-creating-real-problems/}. During the registration phase, user identity information is stored in a Central Database (CD). As CD is compromised, user identity information can be leaked. Also, user identity and passwords are stored on CD for the authentication process, which could be compromised. Considering the flaw in such cases, we aim to utilize decentralized blockchain technology to authenticate and authorize users using privacy-preserving attributes and return control of a user's identity to the user.

Because of its decentralized structure, visibility, traceability, individual control of data in blockchain~\cite{article4,article5}, it has been introduced in a variety of applications, including supply chain~\cite{article6,article7}, healthcare~\cite{article11,article12}, industrial internet of things (IIoT)~\cite{article8,article9,article10}. Miners pack the transaction into a package in a blockchain system, and a consensus algorithm is used to verify the block construction in a network. After verifying a transaction, a legitimate block is added at the end of the longest chain of verified blocks—each node on the blockchain stores a copy of the distributed ledger. Using the blockchain in the identity management system solves traditional schemes that rely on trusted third-party~\cite{article14,article15}. In a standard blockchain system, data is stored as plain text on the network in some schemes~\cite{article16,article17,article18}, violating the privacy-preserving requirement. Our proposed architecture supports attribute-based access control. The blockchain holds re-certification events of the users' identities and the data access and authorization request instead of storing the complete user identity information. These requests stored on the blockchain provide auditability and traceability of the transactions.

In the traditional identity management systems, users give away too much information as the complete identity is shared with the service providers. We want to allow the users to gain control over their identity by choosing what attributes of their identities they want to share to access the services; this can be achieved by attribute-based access control (ABAC) system~\cite{attribute-based}. ABAC is globally recognized and allows or denies user requests based on any user and object's properties and environmental factors that may be more relevant to the rule in question. Our proposed solution allows us to segregate different personas of a user's identity and allow them to decide what attributes and which identities they want to use. Attribute-based access control with OpenID Connect~\cite{sakimura2014openid} ensures that we can utilize existing technologies to ensure fine-grained access control.

We have designed a private blockchain-based system for identity management. Our system suggests a mechanism that would digitize documents such as passports, driving licenses, electricity bills, etc., issued by any government authority or other authority in an immutable and secure manner. Since these authorities already have access to slivers of our identity, we will not be storing these identities on the blockchain. Data owners will still be the custodians of user identity, but users will control how they want to use their identity attributes. Putting personal information on the ledger jeopardizes users' privacy and violates current privacy laws (GDPR, right to be forgotten, etc.) A user's identification traits are also dynamic. Thus, we would associate users' identities on the blockchain with the help of pseudo-anonymous identifiers. The data owners are responsible for authenticating and propagating the users' identities as and when needed. An identity provider could communicate with the consortium of authorities to validate the users' claims regarding their identity. 

\textbf{Our contributions can be summarized as follows:}

\begin{enumerate}
    \item \textbf{We propose a blockchain-based identity management system that enables easy data authentication and data authorization using multiple data sources. Our proposed system emulates real-world conditions where the data owner owns identities. The problem of multiple authentications required by different service providers can be effectively reduced due to our design's federated identity management aspect.}
    \item \textbf{A private blockchain ensures the integrity of data. Even though we do not store the actual identities on the blockchain, we store data access requests and identity attribute re-certification events; this improves the transparency and traceability of the system.}
    \item \textbf{Our proposed system allows users to access different personas of their identity for availing online services based on specific use-cases. We propose an attribute-based identity management scheme that allows fine-grained access control to identity with a trust score associated with the identity. }
    \item \textbf{To ensure user security and confidentiality, we use an identity-based conditional proxy re-encryption scheme that allows transmission of encrypted identities across our system. We also use hierarchical deterministic keys to improve user privacy and increase the traceability and auditability of transactions.}
\end{enumerate}

The proposed system finds its applications in citizen transfers, inter-country service providence, banks, ownership transfer, etc. The generic framework can also be extended to a consortium of banks, hospitals, etc. 

The paper is organized as follows. Section~\ref{sec:related} summarizes related work. Section~\ref{sec:preliminaries} gives the overview of the proposed architecture and the various components utilized in organizing and defining the end-to-end process proposed in this paper. In Section~\ref{sec:methodologies}, we have described the tools and protocols used to draft the system architecture that has been used to ensure the anonymity of users and auditability of transactions. In Section~\ref{sec:system-architecture}, we have defined the communication pattern between user identity providers, service providers, and data owners residing on the blockchain and hence giving the complete system architecture information. The impacts of using this system architecture have been discussed in Section~\ref{sec:discussion}. Section~\ref{conclusion} concludes our work, presenting our limitations and proposing future work.

\section{Related Work}
\label{sec:related}
Earlier, all the user's data was handled by organizations, as the conventional identity management systems mostly depended on third parties. As third parties handle the user's data, there are chances of data being compromised or lost if third parties crumble. Moreover, the organizations can share the data themselves for money or other reasons. The main challenge in centralized systems is transferring users' confidential data from one application to another, as the centralized system is a closed system. There are various proposals given to rectify this problem of centralization. One of the solutions proposed is to use federated identity management \cite{background1,background2,background3,background4}. In federated identity management, a user with a single identity can access many applications with a single registration identity. The problem with these schemes is that the user's data is stored in plain text and controlled by a single organization. Hence, more focus is put on the privacy protection of users' data.

The user-centric identity management \cite{background5,beltran2017user} emphasizes more on the user-oriented paradigm, which allows users to select the data they want to share and, during the authentication process, can present the correct credentials. Laborade et al. \cite{background7} provide a decentralized and user-centric identity management scheme; in this case, the usability is increased as it eliminates user passwords and, with regards to privacy and sovereignty, makes the identity more trustworthy. Singh et al. \cite{background8} try to create a privacy-aware personal data storage (P-PDS) that can take security mindful choices on third parties to get to demand following user inclinations. Nevertheless, the user-centric approach is a weak model. We can take the example of Facebook Connect; here, Facebook is the sole identity provider. Some blockchain identity management schemes are proposed for a decentralized system. In other fields like the financial field, to not reveal the identity of the parties involved in the transactions, decentralized coin-mixing-based methods are suggested \cite{background9,background10,background11}.
 
Similar work in this field has been developed by Hardjono T. and Shrier D.~\cite{hardjono2019core} where the researchers present the argument that the use of MIT OPAL /Enigma for privacy-preserving data sharing offers a better foundation for future internet identities based on information about a user. In a similar technical report by Hardonjo T., et al.~\cite{hardjonoanonymous} the authors address the issue of retaining user anonymity within a permissioned blockchain. The paper presents the ChainAnchor architecture that adds identity and a privacy-preserving layer above the private or the public blockchain. Goyal et al.~\cite{goyal2006attribute} discuss the development of a new cryptosystem called Key-Policy Attribute-Based Encryption (KP-ABE), where ciphertexts are labeled with sets of attributes and private keys are associated with access structures that control which ciphertexts a user can decrypt. Bethencourt A. et al.~\cite{bethencourt2007ciphertext} extended this work in their paper, which describes a system wherein attributes are used to describe a user's credentials. A party encrypting data determines a policy for who can decrypt, unlike previous attribute-based encryption systems that used attributes to describe the encrypted data and built policies into users' keys. 
 
Early work by Paul Dunphy et al.~\cite{8425607} provides an early glimpse of the current strengths and limitations of applying Distributed Ledger Technology (DLT) to Identity Management (IDM) using Cameron's evaluative framework. Xu et al.~\cite{background12} produced the idea of blockchain-based identity management and authentication methods for mobile networks in which users create their SSI identity, and each user will have private and public keys and the user has control over their identity. Gao et al.~\cite{background13}  came up with a blockchain-based privacy protection and identity authentication scheme by using different algorithms like ECDSA encryption and RSA encryption algorithms. We can reduce the storage as well by using these algorithms. This proposal gives access to users to hash their identity and decide whether to store the identity information on a blockchain or not. Faberet al.~\cite{background14} gave a high-level architecture and conceptual design for blockchain-based identity management. This proposal emphasizes giving more control to users of their data. Rathor and Agarwal~\cite{background15} used blockchain in a different area where they created three different modules to give the information to the people looking for a job with the data like experience and education, which helps the organization to verify its employees' records.\cite{background12,background13,background14,background15} can rectify issues like the privacy of user's data and giving the control of user's identity back to a user. However, there is a problem with IP traceability for the IP applications as they need to trace the IP in case of any disputes. Chuxin et al.~\cite{background16} proposed a blockchain-based privacy-preserving and traceability intellectual property (IP) identity management scheme, in which the user's real identity information is processed into multiple shares using improved-Shamir secret sharing, that can reduce storage overhead and achieve privacy protection. Here, user information is stored on the blockchain, thus increasing storage and scalability overheads.

Most of the proposed solutions presented here either store the user's encrypted data on the blockchain or do not account for the auditability of transactions. Also, many of the proposed system architectures assume direct communication between the blockchain and the user. However, this would not be possible in real-world scenarios, leading to remarkably high API response times. Our proposed system builds on the existing federated authentication and authorization systems. This paper proposes an architecture where original data owners are still the custodian of the identity. The user can choose what identities can be used to avail of online services. Our proposed architecture supports attribute-based access control. The blockchain holds re-certification events of the users' identities and the data access and authorization request instead of storing the complete user identity information. These requests stored on the blockchain provide auditability and traceability of the transactions. This would lead to lower resistance to adopting the system than in prior work. The attribute-based identity management system leads to a crucial criterion to minimize identity theft risk.

\section{Preliminaries}
\label{sec:preliminaries}
The identity of an individual is composed of many facets. Every service provider does not necessarily need to know every detail about a user's identity. However, when we use federated identity providers, we give away so much unnecessary information about ourselves; this leads to massive privacy violations, as discussed before. 

The system architecture proposed in this paper has four main components:

\begin{enumerate}
    \item \textbf{Consortium of authorities on a private blockchain}: This could be a consortium of government authorities, credit bureaus, etc. It will also have a querying API server (Communication Server) that will be very tightly coupled to the consortium, allowing an identity provider to query and verify the identity attributes. The communication server is also a node on the blockchain, with some additional responsibilities, including communicating with the outside world.
    \item \textbf{Identity provider (IDP)}: This module deals with user authentication and authorization. It would use OpenID Connect Protocol (OIDC) to implement Single-Sign-On functionality. It will essentially act as an OIDC provider.
    \item \textbf{Service provider (SP)}: This is essentially any third-part service that a user would use. They act as OIDC relying parties.
    \item \textbf{User}: Users can have different privacy-preserving personas, and in our proposed system, users have the liberty to control the persona that they want to use to interact with IDP and hence giving control back to the user to decide the facet they want to use.
\end{enumerate}

The consortium of authorities and the individual would be data owners, whereas the service providers would be data consumers. Our proposed system allows individuals to gain back control of their digital identities. Our system uses decentralized public identifiers and associated descriptor objects to associate user identities on the blockchain. This would ensure that we do not store any Personally Identifiable Information (PII) on the blockchain and always refer to them using pseudo-identifiers. This is critical because a distributed ledger is immutable, meaning anything added to the ledger cannot be changed or removed.
As a result, no personal information should ever be recorded in the ledger. The identity providers would communicate with a server associated with the consortium to get details about the credentials as provided by the user. The identity provider would get user consent to transfer/verify specific attributes of the users' identity to the service provider. We recommend using Open ID Connect since it is the industry standard in federated authentication.

\begin{figure}[h!]
    \centering
    \includegraphics[width=\linewidth,height=\textheight,keepaspectratio]{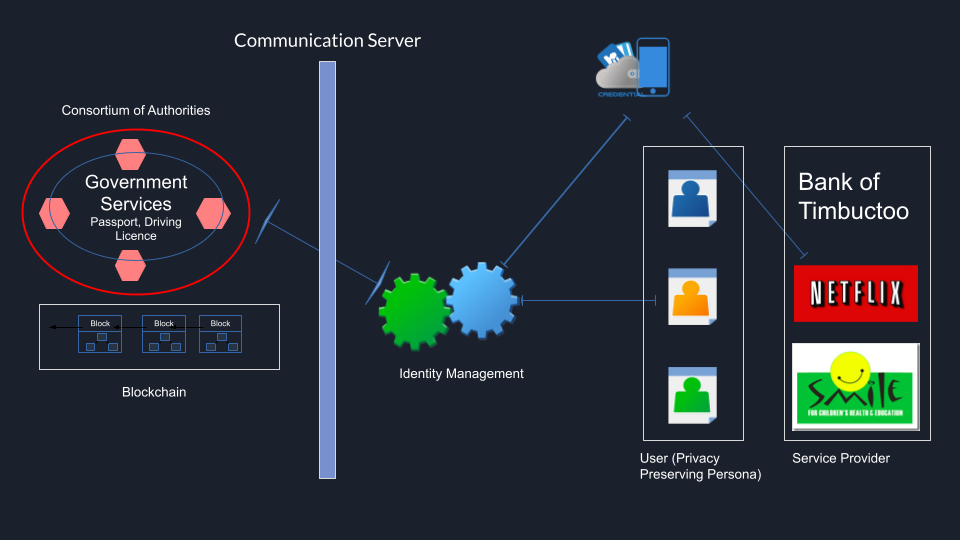}
    \caption{Architecture Overview}
    \label{fig:architecture_overview}
\end{figure}

Figure~\ref{fig:architecture_overview} gives an overview of the proposed architecture. Netflix, Bank of Timbuctoo, etc., are examples of service providers. "Identity Management" is the identity provider in charge of communicating with the consortiums. Our proposed system could have multiple consortiums of identity sources. This would allow the user to control which identity consortium to use depending on the specific use case. For example, a service provider like Netflix might not necessarily need a high trustworthy source of information, so the user could use their social media networks to provide details about specific identity attributes. However, the Bank of Timbuctoo would need very high attestation that the users' identity attributes are correct. In this scenario, the user could verify their identity by the government authorities consortium. This segregation of identity into various slivers gives users more control over their own identity. Additionally, users can decide the privacy-preserving persona they want to use to interact with the identity provider. The blockchain layer stores the re-certification events of the identities and data access and authorization requests, providing auditability and traceability of the transactions.

\section{Tools and Protocols}
\label{sec:methodologies}
This section discusses the different tools and protocols to design our identity management system. We suggest using Hyperledger Fabric, a permissioned-blockchain for storing user transactions. A permissioned blockchain is not publicly accessible and can be accessed only by users with permissions; this provides an additional level of security over permissionless blockchain systems such as Bitcoin or Ethereum, as they require an access control layer. Next, we talk about OpenID Connect, an open and trusted authentication protocol that adds an identity layer to OAuth 2.0. Clients can utilize OIDC to verify an end user's identity utilizing authorization server authentication. We look into the various cryptographic protocols that our design would use to ensure confidentiality of data, transparency of transactions, and guarantee privacy, integrity, and non-repudiation. Lastly, we will look at a mechanism for generating attribute trust scores. An identity provider computes a time-varying confidence score for an asserted attribute and includes it in a security assertion returned to a service provider.

\subsection{Hyperledger Fabric}
Hyperledger Fabric~\cite{androulaki2018hyperledger} is a distributed ledger platform that is open, tested, and enterprise-ready. It is a permissioned ledger with advanced privacy restrictions, which means that only the data you want to be shared is shared among the "permissioned" (known) network participants. Fabric allows establishing decentralized trust in a network of known participants rather than an open network of anonymous participants. The transactions are confidential and only shared with parties that have permission to read those transactions. The transactions are kept private and are only shared with parties who have been granted permission to read them. A ledger resides within the scope of a channel; it can be shared across the entire network (assuming all participants are using the same shared channel) or it can be privatized to only include a limited number of participants. As a result, Fabric is an ideal option for a decentralized but trusted ledger that protects user privacy. 

Hyperledger Fabric's modular architecture divides the transaction processing workflow into three stages: smart contracts, also known as chaincode, consisting of the systems' distributed logic processing and agreement, transaction sequencing, and transaction validation and commitment. This separation has many advantages:

\begin{itemize}
    \item A reduced number of trust levels and verification that keeps the network and processing clutter-free.
    \item Improved network scalability.
    \item Better overall performance.
\end{itemize}

We suggest using Hyperledger Fabric to build a consortium of blockchains; this would allow similar entities to be a part of the same consortium. For example, a consortium of federal government entities (Passport Authority, DMV, etc.) could be separate from a consortium of credit bureaus (Experian, Equifax, etc.). Having separate consortiums of similar entities would allow these entities to open private channels among them as and when needed, thus allowing more accessible communication between them while allowing the user to be wary of these communications.

\subsection{Cryptographic Protocols}
\label{crypto}
This section will discuss the various cryptographic protocols and primitives that our proposed system would use. These protocols ensure data privacy and integrity while also guaranteeing non-repudiation of transactions. Our proposed system uses Hierarchical Deterministic Wallet to generate user keys and Identity Based Conditional Proxy Re-Encryption to securely store and transmit users' personal information throughout the system.

\subsubsection{Hierarchical Deterministic Wallet}
\label{hdw}
Bitcoin and its derivatives use a feature known as Hierarchical Deterministic  Wallets (HD Wallets) that causes your receiving address to change after being used; this is done by creating a ``master" key known as the Extended Private Key and Extended Public Key. This feature enhances the privacy as well as security of the users. The Extended Private Key is the base from which all of your addresses' private keys are derived. Alternatively, in other words: the Extended Private Key is the master key to all the Private Keys belonging to an account. Bitcoin Improvement Protocol 32 (BIP-32)~\footnote{\label{bip-32-footnote} https://github.com/bitcoin/bips/blob/master/bip-0032.mediawiki} describes the specification intended to set a standard for deterministic wallets. BIP-32 uses elliptic curve cryptography using the field and curve parameters defined by secp256k1~\footnote{http://www.secg.org/sec2-v2.pdf}. 

\begin{figure}[h]
    \centering
    \includegraphics[width=0.5\textwidth]{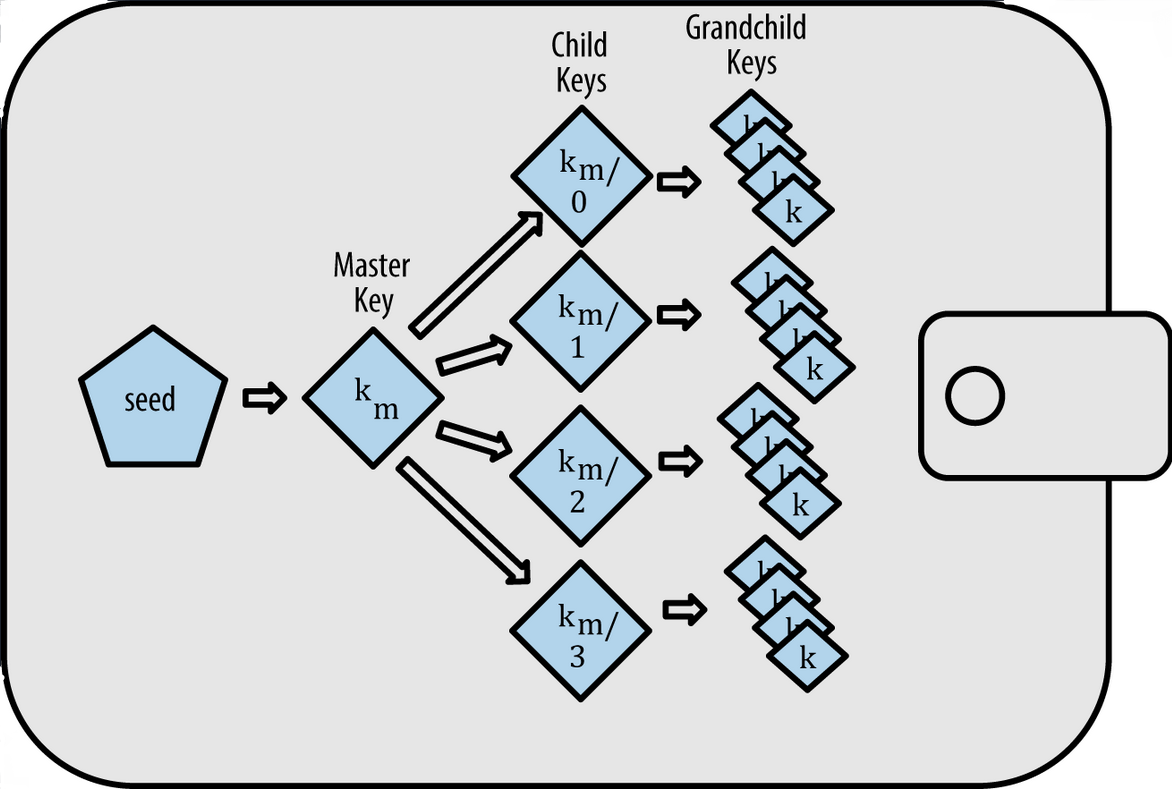}
    \caption{Key Generation Scheme~\cite{josh_2021}}
    \label{fig: hdw}
\end{figure}

Figure~\ref{fig: hdw} showcases how BIP-32 generates keys hierarchically. In our proposed system architecture, the user would hold the master key. The user would then distribute the derived keys as and when needed to relevant authorities/entities in the system. The user would generate two master keys in this entire process: one of them would be associated with all communications between the data owners. In contrast, the other would be associated with authentication with the identity provider. We describe the entire key generation and key distribution process in Section ~\ref{key-generation}.

Since the user would have access to the master key, they would be able to monitor all transactions (in this case, data access authorization approvals). In contrast, the authorities would only be able to monitor transactions from the keys they have been given access. The changing keys have two significant advantages: increased privacy and increased security. By having multiple keys, the user would be able to segregate their identities on the different consortiums of blockchains, enhancing user privacy. Also, since the user has different keys across different data owners, an adversary would need to get multiple private keys to access all the different identities of the user.

\subsubsection{Identity Based Conditional Proxy Re-Encryption}
\label{ibcpre}
Identity-based conditional proxy re-encryption (IBCPRE) is an extension of proxy re-encryption. It provides conditional proxy encryption and extends the proxy re-encryption concept to the identity-based public-key cryptography setup. By using conditional proxy re-encryption, a proxy can re-encrypt a ciphertext using an IBCPRE scheme. However, if a condition is given to the ciphertext and the re-encryption is satisfied, the ciphertext will be well-formed for decryption; this permits fine-grained proxy re-encryption and can be helpful for applications such as safe sharing via encrypted storage.

IBCPRE allows users to choose recipients of a message even after encrypting their messages and uploading them to the server. IBCPRE support end-to-end as well as one-to-many encryption. The IBCPRE's conditional "tag" allows fine-grained access to encrypted messages. The data owner can control the exact set of encrypted communications they want to share with any particular system entity by assigning different tag values to distinct encrypted messages.

Liang et al.~\cite{liang2012cca} was the first to propose a single-hop Identity Based Conditional Proxy Re-Encryption that was proved secure against adaptive conditions and adaptive identity chosen-ciphertext attacks in the standard model. We adopt this scheme for our proposed system and explain it further in Section ~\ref{key-generation}.

\subsection{Trust Score}
\label{trust-score}
Gokhale et al.~\cite{gokhale2021identity} suggested the use of a time-varying trust score for an asserted attribute value. The identity provider would compute a time-varying confidence score for an asserted identity attribute and include it in a security assertion returned to a service provider. The confidence score typically "deteriorates" (i.e., decreases over time). One or more qualifying attribute verification events, on the other hand, may affect the degree to which the score deteriorates.

In our proposed system, we aim to use the identity attribute scoring framework~\cite{gokhale2021identity} to assign trust scores to different attributes. The identity provider would get the verified attribute(s) from the data owner(s). Based on the type of data owners and attributes, the identity provider would generate a trust score and return the said trust score in a security assertion to the service provider. If multiple data owners verify the identity attribute, the trust score could be higher. The identity provider could also profile the service providers and determine the threshold values for an attributes' trust score that the service provider deems acceptable. Based on this profiling, the identity provider could recommend data sources that might be useful for achieving the threshold for a specific attribute. The trust score generation methodology could be proprietary to an identity provider, thus incentivizing multiple identity providers to be a part of the system. 

We explain the necessity of a trust score with a few specific identity attributes:

\begin{itemize}
    \item \textbf{Name}: Governmental data sources would be a more reliable source of information for truthful data about the attribute ``Name" rather than a social media platform or a delivery service platform. However, suppose the service provider does not need a high trust score for this attribute. In that case, the user could choose social media platforms as a source of data information to satisfy the lower threshold of the service provider.
    \item \textbf{Address}: A delivery service provider might have more recent addresses as compared to a government document. Thus, sourcing ``address" from a delivery service provider might result in a higher trust score than a governmental data agency. However, a recent re-certification of this attribute with a governmental agency might result in a better trust score. Thus, it is up to the user to choose relevant data owners to verify their identities based on the specific requirements of the service provider.
\end{itemize}

Our proposed system allows service providers to be more legally compliant. For example, Facebook requires that its users be at least 13 years of age to be able to use their platform~\footnote{https://about.fb.com/news/2021/07/age-verification/}; however, they currently have no way of verifying this. Our system would let them verify the veracity of the ``minimum age" claim without compromising the user's identity.

\subsection{OpenID Connect Protocol}
\label{oidc}
OpenID Connect (OIDC) is an open authentication protocol that adds an identity layer to OAuth 2.0 by profiling and extending it~\cite{sakimura2014openid}. Clients can utilize OIDC to verify an end user's identity utilizing authorization server authentication. By layering OIDC on top of OAuth 2.0, a single framework is created that promises to protect APIs, mobile native apps, and browser apps in a single, unified design. 

Our system would use OpenID Connect for user authentication and identity attribute authorization. OIDC allows service providers to verify the end-user's identity based on the authentication performed by an identity provider and obtain basic profile information about the end-user in an interoperable and REST-like manner. The identity provider obtains the user's basic profile information from the data owners. After obtaining this information, it calculates the trust scores for the requested identity attributes and returns those values with the identity assertions.

\subsection{Key Generation}
\label{key-generation}
This section describes the various keys and secrets being used throughout our system. We will look into the key generation, ownership, and distribution process. We assume that the communication channels between all the entities are secure. We also assume that the user knows the public keys required for IBCPRE of the identity provider and data owners.

\subsubsection{Hierarchical Deterministic Keys}
\label{keygen-hdw}
The user uses the general idea mentioned in BIP-32 to generate two private keys: ``Data Access Key" and ``Data Authorization Key." The Data Access Key is associated with the data owners, and the Data Authorization Key is associated with the identity provider. Having two separate master keys, Data Access Key and Data Authorization Key, allows us to keep the data owners' information separate from the identity providers. The user generates these two keys and keeps them private and safe; leaking a private key would mean a loss of privacy. A ``Child Key Derivation Function" computes a child key given a parent key and an index $i$. Modifying $i$, we can get new child keys~\footnote{See footnote \ref{bip-32-footnote}.}.

\begin{figure}[h!]
    \centering
    \includegraphics[width=\textwidth]{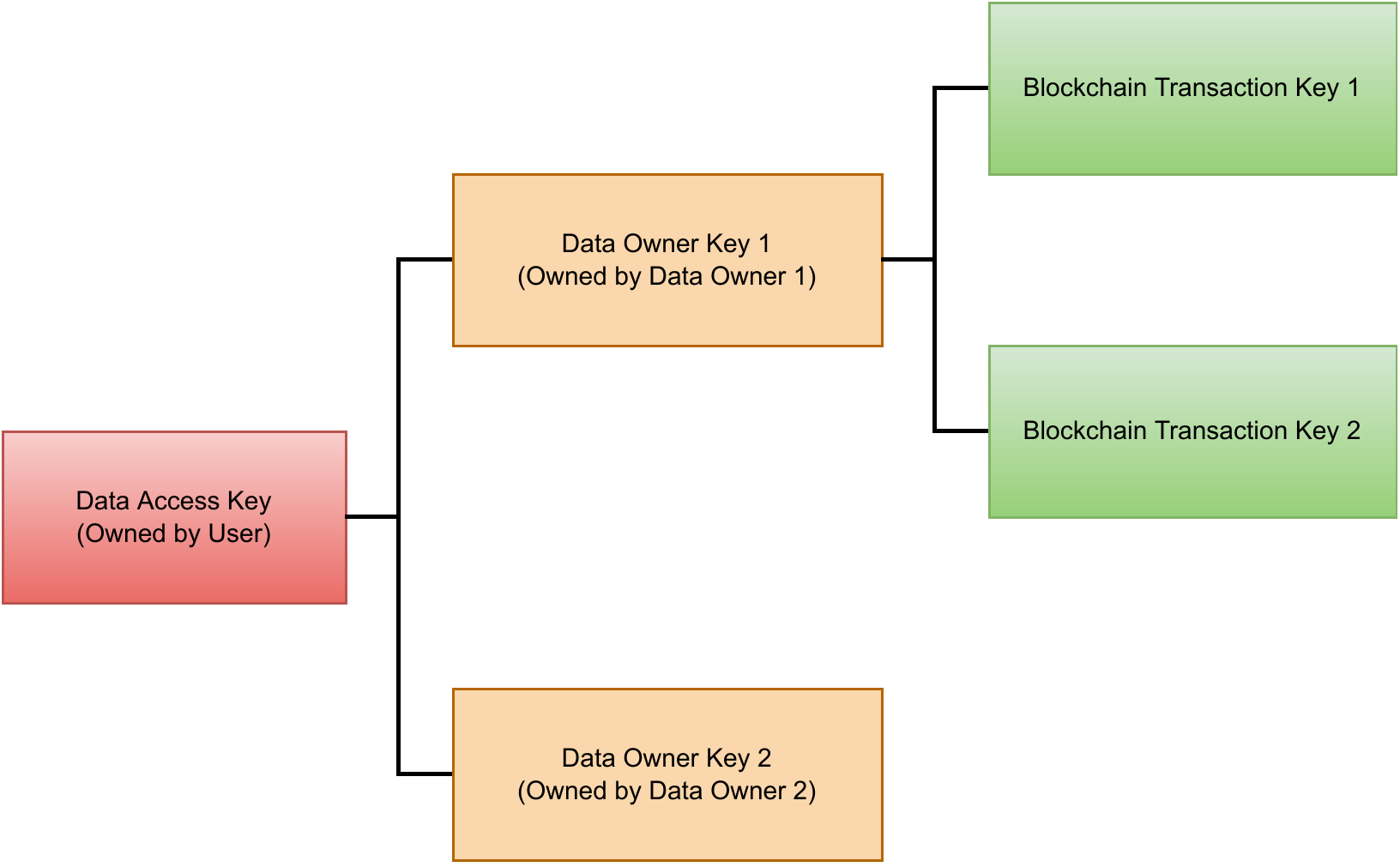}
    \caption{Key Generation Process for Data Access Key}
    \label{fig: key-gen-access}
\end{figure}

The user could generate child private keys for individual data sources from the Data Access Key. These are referred to as ``Data Owner Keys". These data owner keys are then registered with every data owner, either during the creation of the identity or later. The data owners have access to the keys they have and any subsequent keys that the data owners derive from their keys. The data owners can derive further keys, which can then be used to associate transactions about a user on the blockchain. That ensures some level of anonymity on the blockchain. Since these data owner keys are derived from the Data Access Key, the user would have access to all subsequent keys that the data owner derives further. Figure ~\ref{fig: key-gen-access} explains the key generation process of the Data Access Key.

\begin{figure}[h!]
    \centering
    \includegraphics[width=\textwidth]{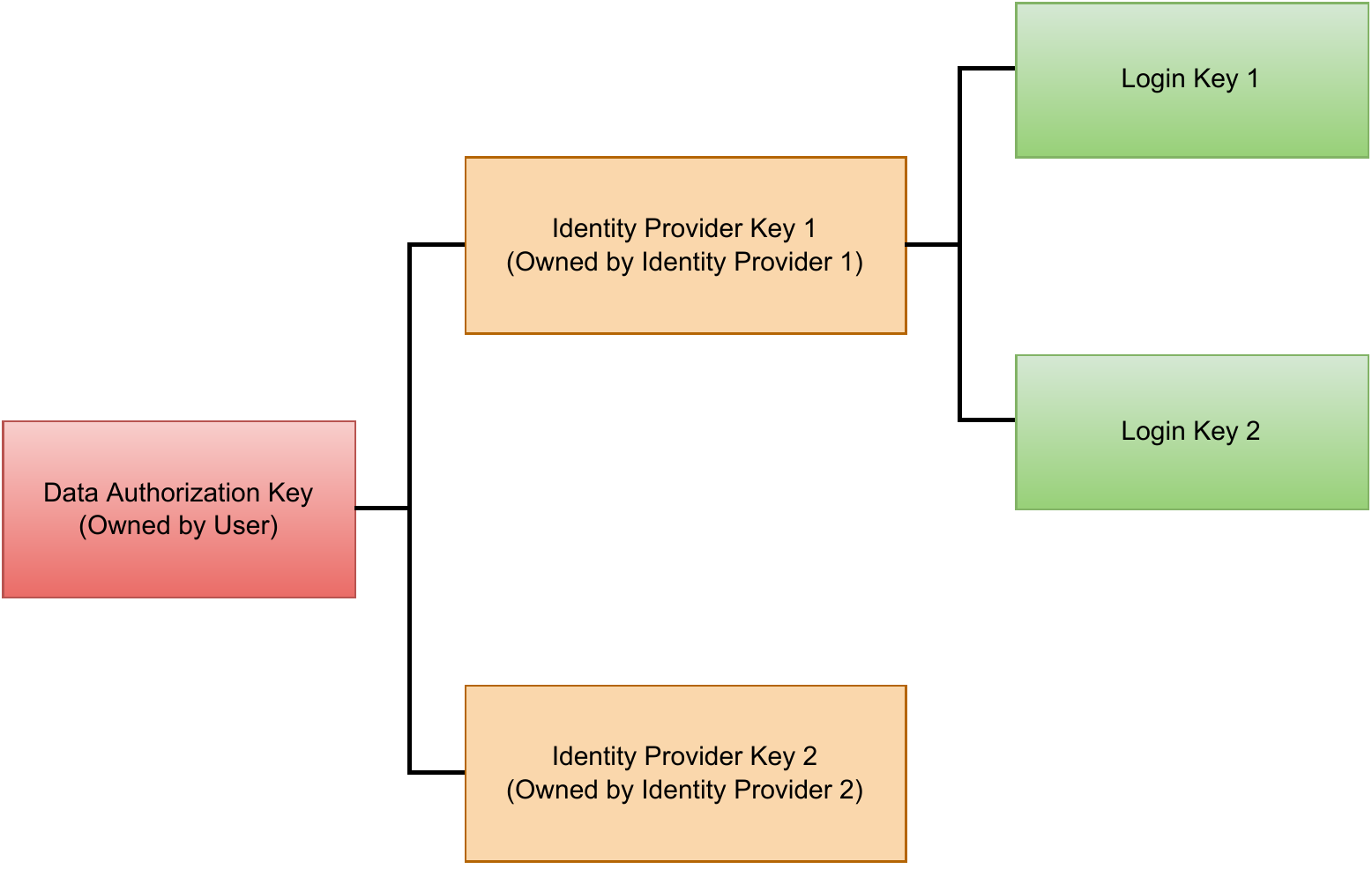}
    \caption{Key Generation Process for Data Authorization Key}
    \label{fig: key-gen-authorization}
\end{figure}

Our system allows for multiple identity providers. We can derive multiple ``Identity Owner Keys" from the Data Authorization Key. These identity owner keys can be used to signup with identity providers and be later used for authentication with the identity provider. Figure ~\ref{fig: key-gen-authorization} depicts the key generation and ownership process. Section ~\ref{system-user-iden-reg} explains the key distribution process.

There are a few shortcomings in the original BIP-32 protocol. Given the master public key and any child private key, an attacker can easily extract the master private key. To tackle this, Gutoski and Stebila~\cite{gutoski2015hierarchical} came up with a better version of HD wallets that is not vulnerable. Their proposed system can handle the leakage of up to $m$ private keys with a master public key size of $O(m)$. In order to improve the security of BIP-32, Das et al.~\cite{das2021exact} came up with a minor modification in the key derivation process of ECDSA. They suggested switching re-randomization in BIP-32 from additive to multiplicative to achieve tighter security without additional costs. They observed that BIP-32 gives roughly 92 bits of security based on their theorems and a conservative choice of parameters; however, the multiplicative version of ECDSA gives 114 bits of security with a similarly efficient scheme. Thus, we recommend using the multiplicative version instead of the additive version for our system.

\subsubsection{Identity Based Conditional Proxy Re-Encryption}
\label{keygen-ibcpre}
We propose to use Liang et al.~\cite{liang2012cca} identity-based proxy re-encryption scheme. Their proposed scheme is collusion resistant, secure against chosen-ciphertext attacks, and supports conditional re-encryption. By associating a condition with the encryption key, a sender can enforce fine-grained re-encryption over their initial ciphertexts. Only the ciphertexts that meet the specified condition can be re-encrypted by the proxy holding the corresponding encryption key.

The user would generate a public key and master private key initially. The user would use the public key of the identity provider and the identity provider's identity to encrypt any data that they would want to store/transmit via the identity provider. Whenever the user wants to share their identities with the data owners for data verification, they would generate a re-encryption key based on the public key of the data owner and the user's identity. This ensures that the identity provider cannot access the identities when transmitting data via the identity provider. The data owners can decrypt the identities using their private keys. Liang et al.~\cite{liang2012cca} explains the process of key generation and provide a security analysis for their proposed scheme. Section~\ref{system-user-login-sp} explains the usage of IBCPRE while the user tries to log in to the service provider.

\section{System Architecture}
\label{sec:system-architecture}
This section explains the communication patterns between the user, identity provider, service provider, and the data owners residing on the blockchain. Our system uses single-sign-on facilities to gain access to online services. We explain the proposed scheme in four phases:

\begin{enumerate}
    \item User registration with data owners.
    \item User interaction with identity provider(s).
    \begin{enumerate}
        \item User registration with identity provider.
        \item User login with identity provider.
    \end{enumerate}
    \item User login with service provider.
    \item User identity registration with identity provider.
\end{enumerate}

We assume that the user's client application has already generated the ``Data Access Key" and ``Data Authorization Key" before proceeding further.

\subsection{User registration with data owners}
\label{system-user-reg-data}
This is the first stage of the process. In this step, the user registers their identity keys with the data owner(s) offline. The user uses their ``Data Access Key" to derive individual data owner keys, as can be seen from Fig. ~\ref{fig: key-gen-access}. The user's client application keeps these details safe. These data owner keys are then registered with respective data owner(s), either during the identity creation or later. These data owner keys will serve as the pseudo-identifier for the individual on the blockchain. This would prevent other consortium members on the blockchain from co-relating these identifiers with the user since every data owner will have their separate data owner key.

\subsection{User interaction with identity provider(s)}
\label{system-user-inter}
This section describes user interaction with the identity provider. We explain the user registration and sign-in process.

\subsubsection{User registration with identity provider}
\label{system-user-reg-idp}
\begin{figure}[h!]
    \centering
     \includegraphics[width=\textwidth]{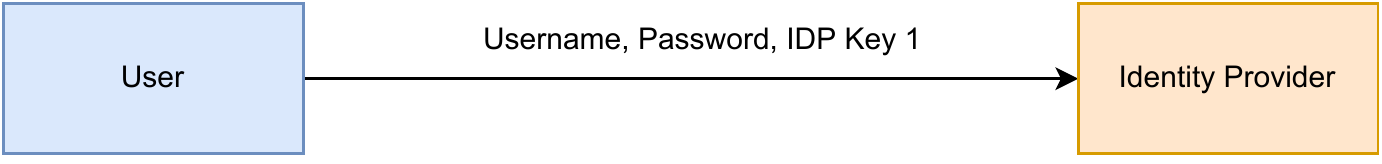}
    \caption{User Registration with IDP}
    \label{fig: user-reg-idp}
\end{figure}

 We propose a multi-factor authentication system for our system design. Figure ~\ref{fig: user-reg-idp} details the user signup process with an identity provider. The user generates an identity provider key from the "Data Authorization Key," as shown in Fig.~\ref{fig: key-gen-authorization}. The user chooses a \textit{username}, \textit{password}, and the recently generated \textit{identity provider key} as signup parameters with the identity provider. The user also signs up with a TOTP-based~\cite{m2011totp} 2FA (Two-factor authentication) authenticator app for added security.

We recommend using multi-factor authentication methods over TOTP-based authentication methods since there have been multiple incidents where the latter has proved to be not secure enough~\footnote{https://www.theverge.com/2019/8/14/20805194/suprema-biostar-2-security-system-hack-breach-biometric-info-personal-data}~\footnote{https://www.varonis.com/blog/box-mfa-bypass-totp}. We also recommend following NIST's password policy guidelines~\cite{burr2011sp}.

\subsubsection{User login with identity provider}
\label{system-user-login-idp} 
\begin{figure}[h!]
    \centering
     \includegraphics[width=\linewidth,keepaspectratio]{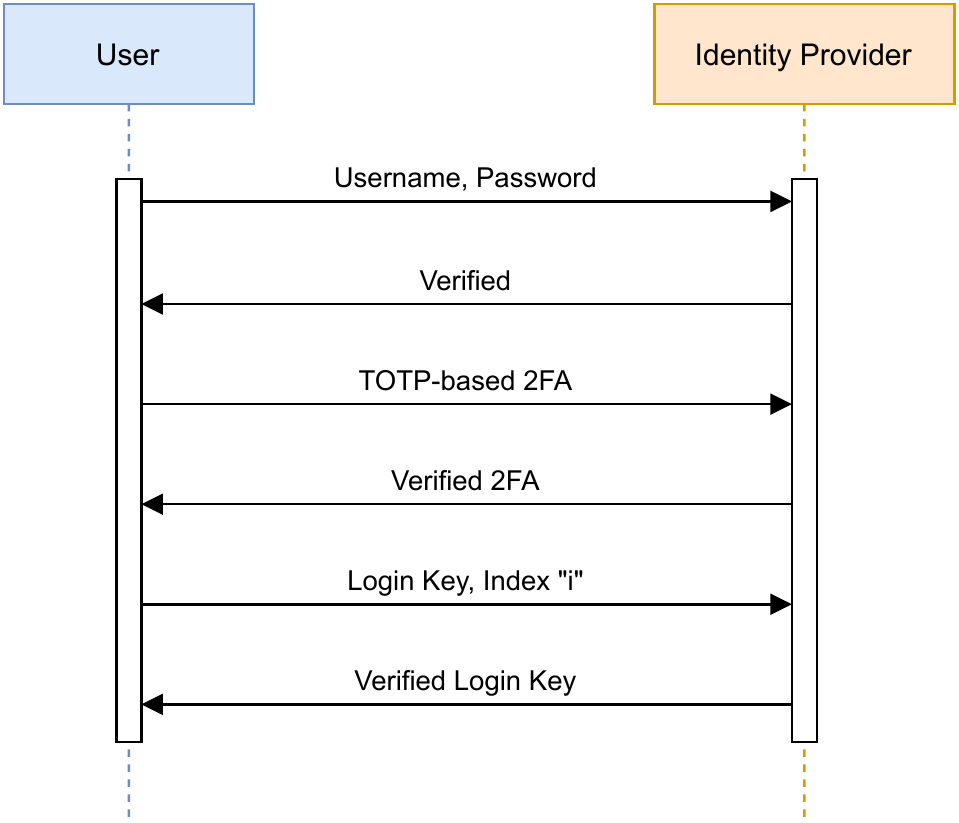}
    \caption{User Login with IDP}
    \label{fig: user-login-idp}
\end{figure}

This section describes the user login process with the identity provider(s). Figure ~\ref{fig: user-login-idp} can be used as a reference to understand this process. There are multiple steps in this phase:
\begin{enumerate}
    \item The user enters the \textit{username} and \textit{password}. If verified, it moves on to the TOTP-based 2FA screen. If this fails, the user is asked to re-enter the details.
    \item Once username and password are verified, the user verifies the 2FA code. If approved, we move to the next step; else, we redirect the user to the failed login state.
    \item Once the 2FA code is verified, the user generates a login key from the identity provider key of the current identity provider key that the user holds. The user generates a login key by modifying the index $i$ in the hierarchical deterministic key generation process. The user sends the login key with the index $i$ to the identity provider.
    \item The identity provider has the identity provider key of the user. It fetches that and uses the input index $i$ received by the user. If the generated key matches the login key that the identity provider got as input from the user, we can safely say that the user authentication was successful.
\end{enumerate}
This multi-step authentication process adds friction to the sign-in process but ensures security. 

\subsection{User login with service provider}
\label{system-user-login-sp}
\begin{figure}[h!]
    \centering
    \includegraphics[width=\textwidth]{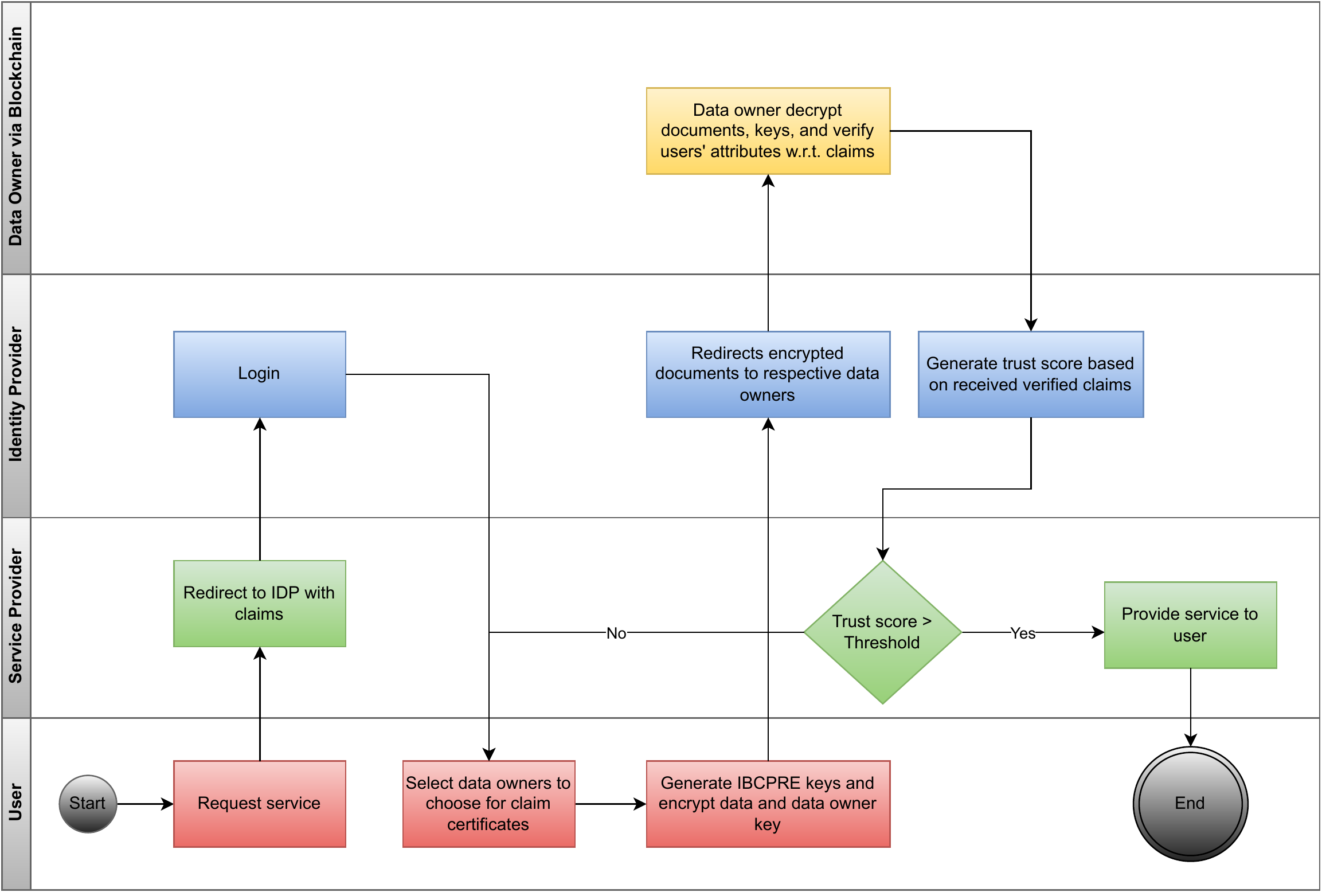}
    \caption{User Login with Service Provider}
    \label{fig: user-login-sp}
\end{figure}

Our system uses OpenID Connect for federated authentication and authorization. All communication between the identity provider and the consortium of data owners is through the communication server of the consortium. Figure ~\ref{fig: user-login-sp} can be used as a reference.

\begin{enumerate}
    \item A user can select the identity provider when they try to log in to a service provider. A user attempting to log in to a service using some IDP is redirected to that IDP with the required claims whenever they wish to use it. 
    \item The user logs in to the IDP. The user can select specific data owners that they want to use to certify certain attributes of their identity. The user can choose relevant data owners based on the service they are trying to access and the required attribute claims. If, for example, a user needs their date of birth for their social security number, they can use their passport as the identity source since it is a better source of truth for the said attribute. However, if a service like an email provider needs a user's name, users can use their social media identities to get those attribute claims. Choosing the data owners depends on the kinds of service the user is trying to avail.
    \item The user uses their secret IBCPRE key to encrypt the data identity and the data owner key. Next, they send these encrypted documents to the identity provider. We need to note that even the identity provider will not be able to access these documents at this stage since they are encrypted.
    \item The user uses the data owner's public key to generate a re-encryption key to re-encrypt the data owner key and identity. These encrypted objects are sent to the data owners, where they decrypt these objects using their secret IBCPRE keys and the list of required attributes.
    \item The data owner decrypts the objects, fetches the identities stored on their systems against the received data owner key and verifies information on their system with the input identity document from the user. If these documents match, the data owner sends a green signal to the identity provider about the veracity of the attributes. It also sends information about the latest re-certification event for the said attribute from the blockchain. 
    \item The data owner adds a new transaction on the blockchain detailing the data access request. The blockchain transaction holds timing information about the data access request, identity provider details, service provider details, and the requested attribute details. The data owner generates a new blockchain transaction key for the user using the data owner key of the user for the above transaction.
    \item Once the identity provider receives the green signal from the data owners, the user generates a re-encryption key for the identity provider. The identity provider can decrypt the already stored documents. Since we are only generating re-encryption keys (IBCPRE) for the data owner and the identity provider, the documents sent to the data owner are the same as being used by the identity provider.
    \item Based on the decrypted documents, the identity provider can calculate a trust score for the required attributes. The identity provider sends over the asserted attribute values and the trust score for the said attribute back to the service provider.
    \item If the trust score is greater than the threshold set by the service provider for a threshold, the user is granted access to the service. Else, they will have to choose a different data owner.
\end{enumerate}

\subsection{User identity registration with identity provider}
\label{system-user-iden-reg}
When the user tries to log in to the service provider via the identity provider, the user has to furnish their data documents every single time. That becomes a hassle and adds friction to the entire process. This section proposes a secure way to store encrypted documents with the identity provider. However, we do not recommend using this method since we have to trust the identity provider to some extent here. Despite the apparent security concerns, this method reduces the time required to verify the identity of a user and reduces the communication overhead with the data owners consortium.

\begin{enumerate}
    \item A majority of the steps remain the same as mentioned in Section ~\ref{system-user-login-sp}. The user logs in to the IDP, selects data sources, and tries to verify their identities with the respective data owners.
    \item If the user's identity document is verified, the data owner sends a green signal to the identity provider. However, no information about re-certification events is sent since this was not requested.
    \item Now, the identity provider stores the verified but encrypted document in their systems. The user can use this every time they want to use these identities to avail of online services.
    \item Whenever a user would try to use their identities stored at the identity provider, the user would generate a re-encryption key using the identity provider's IBCPRE public key. The identity provider would decrypt the stored document and verify the identity attribute claims being requested by the service provider.
\end{enumerate}

One apparent disadvantage to this method is that the trust score of an attribute might be slightly lower since the identity provider has no information about the re-certification events. Also, the user might not be able to track their data access requests as they could using the blockchain. Nevertheless, this does provide a possible business opportunity for the identity providers to lure users into using their services since they might provide additional features such as tracking data access requests. Essentially, it is up to the user to decide what identity provider they want to use.

\section{Discussion}
\label{sec:discussion}
In this section, we analyze the security of our proposed scheme and discuss our contributions with respect to the existing systems.

\subsection{Identity Security}
Firstly, the users generated the Data Access Key and Data Authorization Key and stored them locally. Thus, no adversary can obtain them without physical access to the user's device. Secondly, the user needs to provide their identity to the identity provider only when trying to access online services. The identities are encrypted using IBCPRE, and the user maintains the encryption keys. Thus, not even the identity provider has access to the documents unless the user generates the re-encryption keys for the identity provider. The Data Access Key is also encrypted using the IBCPRE scheme when in transit, thus disallowing the identity provider from ever gaining access to it. Lastly, Data Access Key and Data Authorization Key are separate and are used for different use-cases altogether, thus, providing the much-needed barrier between data authentication and data authorization.

\subsection{Data Confidentiality and Privacy}
Our proposed system uses state-of-the-art cryptographic algorithms to ensure the confidentiality of data. Our IBCPRE scheme uses AES-512 to encrypt the documents/information in transit. Also, a user creates a different data owner key for every data owner, ensuring the segregation of information between the data owners. Our hierarchical deterministic key generation scheme allows users to have a separate transaction key on the blockchain every time a new transaction is committed, thus reducing the chances of data correlation if an adversary gains unauthorized access to the blockchain. The blockchain does not store any private information about a user, just metadata about data access requests and re-certification events.

\subsection{Data Transparency}
The blockchain stores only the identity attributes' data access requests and re-certification events. Since the transaction keys used for these transactions are derived from a single parent key, the data owner can trace all the transactions belonging to the same user. This assures traceability and auditability of transactions as and when needed. Also, the user can trace back their data access requests since they hold the master Data Access Key; the user can trace transactions on the blockchain for all data owners.

\subsection{Transaction Integrity and Non-Repudiation}
Firstly, using a trusted, permissioned blockchain ensures that all consortium members of a private blockchain trust each other. Since these transactions on the blockchain are immutable and permanent, they cannot be modified or deleted, thus ensuring the system's integrity. Secondly, only the user holds the Data Access Key and can thus initiate and finalize a data access request. This ensures non-repudiation of the identity.

\subsection{Authentication and Authorization}
In our proposed scheme, we follow a multi-factor authentication strategy. Section ~\ref{system-user-login-idp} explains how we maintain security during the login phase. The identity provider has no idea about the user's identity but only serves as a trusted proxy between the user and the service provider. It relies on the data owner to verify the identity of the users. Our proposed system suggests using attribute-based access control policies to limit data leakage to the service provider. The service provider could request as many attributes as they want; however, it is up to the user to authorize/unauthorize what attribute information needs to be propagated to the service provider. The service provider can make specific attribute requirements mandatory and limit access to complete service functionality based on the approved attributes it receives from the user. For example, suppose the user does not wish to disclose their age to a media streaming platform. The media streaming platform could disallow users from viewing adult content on their platform. Our platform establishes that even though the user has control over their identity, they would still have to face the consequences of not allowing access to the required attributes; however, at least the user has a choice.

\subsection{Scalability and Availability}
Our proposed scheme stores only re-certification events and data access requests on the blockchain and not the complete identity~\cite{background16}. Hence, this allows us not to overload the blockchain, thus improving the system's scalability. In the real-world, data owners are regarded as data custodians. For example, the passport of a citizen is still the property of that country's government and not the user. Thus, we extend this same philosophy over the internet and do not focus on self-sovereign identities like a few previous studies~\cite{selfsovereign1,selfsovereign2}. The system can support large-scale user identity management and authentication by supporting multiple consortiums of data owners.

One limiting factor in the proposed scheme is the availability of data sources when needed. If the data sources are offline, the user might be blocked. We can overcome this by letting users upload their identities on the identity provider (though not recommended). The system could also be modified to allow the identity provider to generate a lower trust score if a data source is unavailable if they have the user's consent.

\section{Conclusion and Future Work}
\label{conclusion}
This paper proposes a blockchain-based, privacy-preserving user identity management system. In our proposed identity management protocol, users can use their existing identities to access services using hierarchical deterministic keys to ensure access to their data and a proxy re-encryption scheme to store and transmit data securely. Our system uses existing single sign-on facilities with attribute-based access control for efficient, fine-grained user authentication and authorization. Furthermore, the user can also trace their authentication and data access requests as and when needed using a secure, permissioned blockchain. Using multiple data owners allows users to maintain and control different slivers of their identity. An identity provider assigns a trust score for every asserted attribute based on the data sources and various attribute re-certification events. This trust score allows service providers to control or limit access to a particular service aspect based on their internal thresholds for attributes. Our proposed identity management scheme gives users control over their data, allowing them to access online services securely.

To that end, there is room for improvement. This study is one of the initial stages toward the implementation of a fully functional privacy-first federated identity management system. In order to understand and utilize the system, the interface will need to abstract away most of the blockchain and identity provider interactions. As a result, more fieldwork in usability testing will be required. We also need to do some performance testing to see if this system architecture is feasible in the real world. While there are cryptographic guarantees for the privacy of the transaction when it is stored, there is always the risk of some data leakage (e.g., by usage, frequent updates, etc.) that might worry policymakers. As a result, further security testing would be required. In the future, we want to resolve these concerns.

\bibliographystyle{splncs04}
\bibliography{bibliography}

\end{document}